\documentstyle[aps,multicol,pre,epsfig]{revtex}

\newcommand{\beq}{\begin{equation}}
\newcommand{\eeq}{\end{equation}}
\newcommand{\ba}{\begin{array}}
\newcommand{\ea}{\end{array}}
\newcommand{\bean}{\begin{eqnarray*}}
\newcommand{\eean}{\end{eqnarray}}
\newcommand{\bea}{\begin{eqnarray}}
\newcommand{\eea}{\end{eqnarray}}
\newcommand{\bc}{\begin{center}}
\newcommand{\ec}{\end{center}}
\newcommand{\bt}{\begin{table}}
\newcommand{\et}{\end{table}}

\newcommand{\bpsi}{{\bf \psi}}
\newcommand{\la}[1]{\label{#1}}

\newcommand{\no}{\noindent}

\newcommand{\rf}[1]{(\ref{#1})}
\newcommand{\beqno}{\begin{displaymath}}
\newcommand{\eeqno}{\end{displaymath}}

\newcommand{\been}{\begin{enumerate}}
\newcommand{\een}{\end{enumerate}}

\newcommand{\sn}{{\rm sn}}
\newcommand{\cn}{{\rm cn}}

\newlength{\myheight}
\newlength{\mylength}

\newcounter{saveeqn}

\newtheorem{example}{Example}

\begin{document}

\begin{title}
{\bf Linearly Coupled Bose-Einstein Condensates: From Rabi Oscillations 
and Quasi-Periodic Solutions to Oscillating Domain Walls and Spiral Waves}
\end{title}
\author{B. Deconinck$^{1}$,  P.G. Kevrekidis$^{2}$, H.E. Nistazakis$^{3}$
and D.J. Frantzeskakis$^{3}$}
\address{
$^1$ Department of Applied Mathematics, University of Washington, 
Seattle, WA 98195, USA \\
$^2$ Department of Mathematics and Statistics, University of 
Massachusetts, Amherst MA 01003-4515, USA \\
$^3$ Department of Physics, University of Athens, Panepistimiopolis, 
Zografos, Athens 15784, Greece }

\maketitle

\begin{abstract}
In this paper, an exact unitary transformation is examined that allows for 
the construction of solutions of coupled nonlinear Schr{\"o}dinger equations 
with additional linear field coupling, from solutions of the 
problem where this linear coupling is absent.
The most general case where the transformation is applicable is identified. 
We then focus on the most important special case, namely 
the well-known Manakov system, which is known to be relevant for applications 
in Bose-Einstein condensates consisting of different hyperfine states of 
$^{87}$Rb. In essence, the transformation constitutes a distributed, 
nonlinear as well as multi-component generalization of the Rabi oscillations 
between two-level atomic systems. It is used here to derive a host of
periodic and quasi-periodic solutions including temporally oscillating
domain walls and spiral waves.
\end{abstract}

\vspace{2mm}

\begin{multicols}{2}

{\it Introduction.} The recent progress in experimental and theoretical 
studies of Bose-Einstein 
condensates (BECs) \cite{review} has made solitary matter waves physically
relevant objects. One-dimensional (1D) dark \cite{dark} and bright \cite
{hulet} solitons have been observed in recent experiments. On the other hand,
optical solitons have a time-honored history as fundamental nonlinear
excitations in optical fibers and waveguides
(see, e.g., the recent reviews \cite{buryak,kivpr}).

A very relevant generalization of this class of physical systems
and of the solitary waves they can support concerns the
case of multiple coupled components. There has recently been
a considerable volume of work relevant to the properties
of coupled BECs ranging from the ground state 
solutions \cite{shenoy,esry} to the small-amplitude excitations 
\cite{excit} of the order parameters. Furthermore, the formation of 
various structures including domain walls \cite{Marek,haelt}, 
bound dark-dark \cite{obsantos}, dark-bright 
\cite{anglin}, dark-antidark, dark-gray, bright-antidark and 
bright-gray soliton complexes \cite{epjd}, as well as spatially periodic 
states \cite{decon} was also predicted. On the other hand, 
experimental results have been reported for mixtures of different 
spin states of $^{87}$Rb \cite{myatt} and mixed condensates
\cite{dsh,stamp}. It is relevant to mention the efforts into the
realization of two-component BECs from different atomic species, 
such as $^{41}$K--$^{87}$Rb \cite{KRb} and 
$^{7}$Li--$^{133}$Cs \cite{LiCs}. 

Typically, the relevant model for two coupled BECs involves two 
nonlinearly coupled Gross-Pitaevskii (GP) equations. However, in experiments 
with a radio-frequency (or an electric field) coupling two separate hyperfine 
states \cite{myatt,hall}, the relevant model involves
also a linear coupling between the wavefunctions. The governing normalized 
equations are then of the form:
\begin{eqnarray}
i \psi_{1t}=\left[-\frac{1}{2} \Delta+V+ a_{11}|\psi_{1}|^{2}+a_{12}|\psi _{2}|^{2}\right]\psi_{1}+ \alpha \psi_{2},\label{lceqa} \\
i \psi_{2t}=\left[-\frac{1}{2} \Delta+V+ a_{12}|\psi _{1}|^{2}+
a_{22}|\psi _{2}|^{2}\right]\psi_{2}+ \alpha \psi_{1},
\label{lceqb}
\end{eqnarray}
where $V \equiv V(r)$ is the relevant potential, typically consisting
of a magnetic trap and/or an optical lattice \cite{review,reviewus}, while 
$\psi_j$'s represent the condensate wavefunctions. The intra- and 
inter-species interactions are characterized by the coefficients 
$a_{jj}$ ($j=1,2$) and $a_{12}$ respectively, 
while  $\alpha$ denotes the strength of the 
radio-frequency (or electric field) coupling.
Note that Eqs. (\ref{lceqa})-(\ref{lceqb}), 
combining both linear and nonlinear 
couplings, occur in fiber optics as well: In that case, $\psi_j$'s are 
two coupled electric field envelopes of the same wavelength but of different 
polarizations and the linear coupling is generated either by a twist 
applied to the fiber in the case of two linear 
polarizations, or by an elliptic deformation of the fiber's core in the case 
of circular polarizations \cite{old,old1} (linear coupling is impossible 
when considering waves of 
different wavelengths). Another optical model, with only linear coupling 
between two 
modes, applies to nonlinear fiber couplers \cite{jen} or dual-core 
nonlinear fibers 
(see, e.g., \cite{UNSW}).
In the context of BECs, this coupling has been recently examined for 
extended wave solutions in \cite{porter}. 

In the present work, we aim to study a unitary transformation in the 
context of 
Eqs. (\ref{lceqa})-(\ref{lceqb}) that completely absorbs the linear coupling
between the components into an oscillatory temporal dependence. We illustrate
the value of this transformation in a two-fold way: on the one hand, we use it
to understand the role of the linear coupling between the components as a means
of creating Rabi oscillations between the matter present in the two components
(e.g., hyperfine states) \cite{rabi} and their analog in power oscillations
between polarizations in optical systems. On the other hand, we use it to
construct {\it exact} time-periodic solutions of such linearly coupled
nonlinear Schr{\"o}dinger equations. This way, we identify time-periodic
Thomas-Fermi clouds and extended waves in one spatial dimension, as well as
sloshing domain walls and vortices in two dimensions. Since the transformation
is exact only in the so-called Manakov case of $a_{11}=a_{22}=a_{12}$
\cite{manakov},
we 
investigate numerically what happens to the constructed time-periodic solutions
in cases (relevant to $^{87}$Rb experiments) where $a_{11} \neq a_{22} \neq
a_{12}$.
Finally, we generalize the transformation to show that analogous constructions
are feasible with a higher number of components. Our focus here is in
illustrating the generality and applicability of the transformation to a very
broad host of settings 
(multiple component, as well as higher dimensional cases, also in the presence 
of external potentials). We should note that special cases of this 
transformation have previously
been identified in optics (in the integrable case of the two-component, 
one-dimensional
Manakov solitons in the absence of a potential \cite{old1}), as well as in a 
different format in BECs [in two-component, 1D condensates in the presence of 
a periodic potential \cite{bdk}, where solutions were constructed as unitary 
transformations of stationary solutions of the equations (\ref{eqn:nls}) 
below. There the nonstationarity was introduced 
by the mixing of two stationary solutions that were not in phase.]

{\it Analytical Results.} 
Assuming that $a_{11}=a_{22} \equiv g$ and $a_{12} \equiv h$,
Eqs. (\ref{lceqa},\ref{lceqb}) take the following matrix form:
\beq\la{eqn:stark}
i {\bpsi}_t -\alpha P \bpsi =
-\frac{1}{2}\Delta {\bpsi}+({\bpsi}^\dagger G \bpsi)\bpsi+V({\bf x})\bpsi,
\eeq
where
\beq
G=\left(
\ba{cc}
g&0\\
0&h
\ea
\right), ~~ 
P=\left(
\ba{cc}
0&1\\
1&0
\ea
\right).
\eeq
Furthermore, in the special case $h=g$ (hence $G=g I$, with $I$ being the identity matrix), we may 
consider Eq. (\ref{eqn:stark}) as a homogeneous equation (LHS) with an 
inhomogeneous part (RHS); in this case, the solution of the homogeneous equation is given by 
\beq\la{unit}
\bpsi=U(t)\bpsi_0=e^{-i \alpha P t}\bpsi_0=\left(
\ba{cc}
\cos(\alpha t)& -i \sin(\alpha t)\\
-i \sin(\alpha t) & \cos(\alpha t)
\ea
\right)\bpsi_0. 
\eeq
Using a variations of parameters approach, we substitute this in the 
equation for $\bpsi$. This results in: 
\beq\la{eqn:nls}
i\bpsi_{0t}=-\frac{1}{2} \Delta \bpsi_{0}+(\bpsi_0^\dagger G \bpsi_0)\bpsi_0+
V({\bf x})\bpsi_0,
\eeq
i.e., the same equation as \rf{eqn:stark}, but without the electric field 
coupling terms. As mentioned above, large classes of exact  stationary and
nonstationary solutions to this equation were  constructed in
\cite{Marek,haelt,obsantos,anglin,epjd,decon,bdk}.  Those solutions can now
be used to construct exact solutions to \rf{eqn:stark}. Importantly, {\it
stable} such solutions, are also relevant to the more general model Eqs.
(\ref{lceqa})-(\ref{lceqb}), and in particular to the BEC experiments with
$^{87}$Rb \cite{dsh} or $^{23}$Na \cite{stamp}: in that case, the deviations of
the values of the nonlinear coefficients $a_{jk}$ from $1$ are typically on the
order of $3\%$. Such a small difference acts as a small perturbation
and does not alter the stability of the solutions 
(see also the numerical results below).
Note that, typically, these solutions will be nonstationary and, in particular,
time-periodic.

The success of this simple variation of parameters method can be phrased 
differently in physical terms: the equation \rf{eqn:nls} is invariant under 
unitary transformations, as it is an additive combination of the linear 
Schr\"odinger equation and a nonlinear term with a matrix $G$ that commutes 
with any matrix. Then the extra linear coupling terms can be removed by using 
a time-dependent unitary transformation, which does not affect the other 
terms of the equation. 
Thus, the solutions of \rf{eqn:stark} can be thought of as rotating 
in ``spin space''. 




We should note that more general nonlinear coupling matrices 
$G$ commute with 
$U$. In particular:
\beq
\tilde{G}=\left(\ba{cc}0&1\\1&0\ea\right),
\eeq
also commutes with $U$. 
The relevant dynamical equations, however, containing the term
$(\psi_0^\dagger \tilde{G} \psi_0)\psi_0$ result in nonlinearities
of the form $|\psi_1|^2 \psi_2$ and $\psi_1^2 \psi_2^{\star}$ in 
the dynamical equation for $\psi_1$ and hence seem less physically
relevant.

The density of the different components $n_i=|\psi_i|^2$ ($i=1,2$) is given by 
\bea\la{eqn:dens1}
n_1 &=& \cos^2(\alpha t)|\psi_{01}|^2+\sin^2(\alpha t)|\psi_{02}|^2
\nonumber
\\
    &+& |\psi_{01}||\psi_{02}|\sin(\theta_2-\theta_1)\sin(2\alpha t),\\ \la{eqn:dens2}
n_2 &=& \cos^2(\alpha t)|\psi_{02}|^2+\sin^2(\alpha t)|\psi_{01}|^2
\nonumber
\\
&+& |\psi_{01}||\psi_{02}|\sin(\theta_1-\theta_2)\sin(2\alpha t),
\eea
where $\psi_{01}=|\psi_{01}|e^{i\theta_1}$, 
$\psi_{02}=|\psi_{02}|e^{i\theta_2}$. Note that if 
$\psi_{01}$ and $\psi_{02}$ are time independent, then 
$n_1+n_2$  does not depend on time. Thus it is a {\it local} constant of the 
motion (as opposed to the global constant $\int (n_1+n_2) dx$ which is 
conserved
for {\it any} G). This is equivalent to the statement that the nonlinear 
term in  \rf{eqn:stark} is invariant under the unitary transformation given 
by $U(t)$. Notice also that the existence of stationary solutions 
of Eq. (\ref{eqn:stark}) requires the expressions
(\ref{eqn:dens1})-(\ref{eqn:dens2}) to be independent of $t$, which only 
happens when both $\psi_{10}$ and $\psi_{20}$ are stationary and equal.
Furthermore, clearly the above unitary transformation illustrates
that the linear coupling does not affect the integrable nature of
the 1D Manakov model (for $V(x)=0$).

{\it Numerical Results.} We now turn to the practical usefulness of
the transformation
i.e., constructing time-periodic solutions
of Eq. (\ref{eqn:stark}) from stationary solutions of Eq. (\ref{eqn:nls}), 
as well as quasi-periodic solutions of the former from periodic ones of
the latter (i.e., the transformation always inserts
an additional frequency in the time-dependence of the solution).
While one can follow this path also in the absence of the potential
for the known, exact solutions of the Manakov model, we will focus herein
on the case {\it with} the potential, which is more relevant to BECs 
\cite{myatt,dsh,stamp,hall}.

\begin{figure}[tbp]
\centering 
\includegraphics[width=2.5 in]{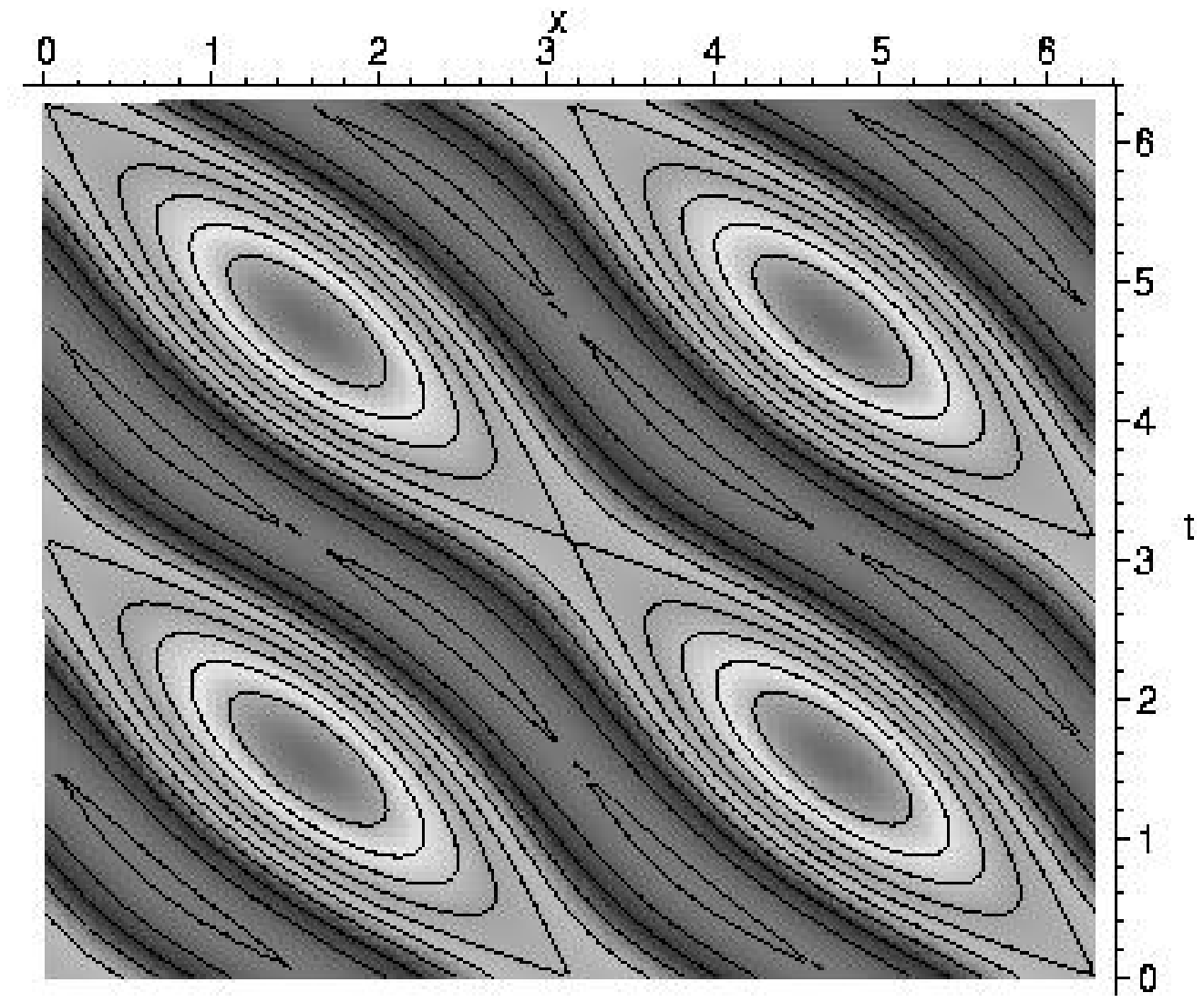}
\centering
\includegraphics[width=2.5in]{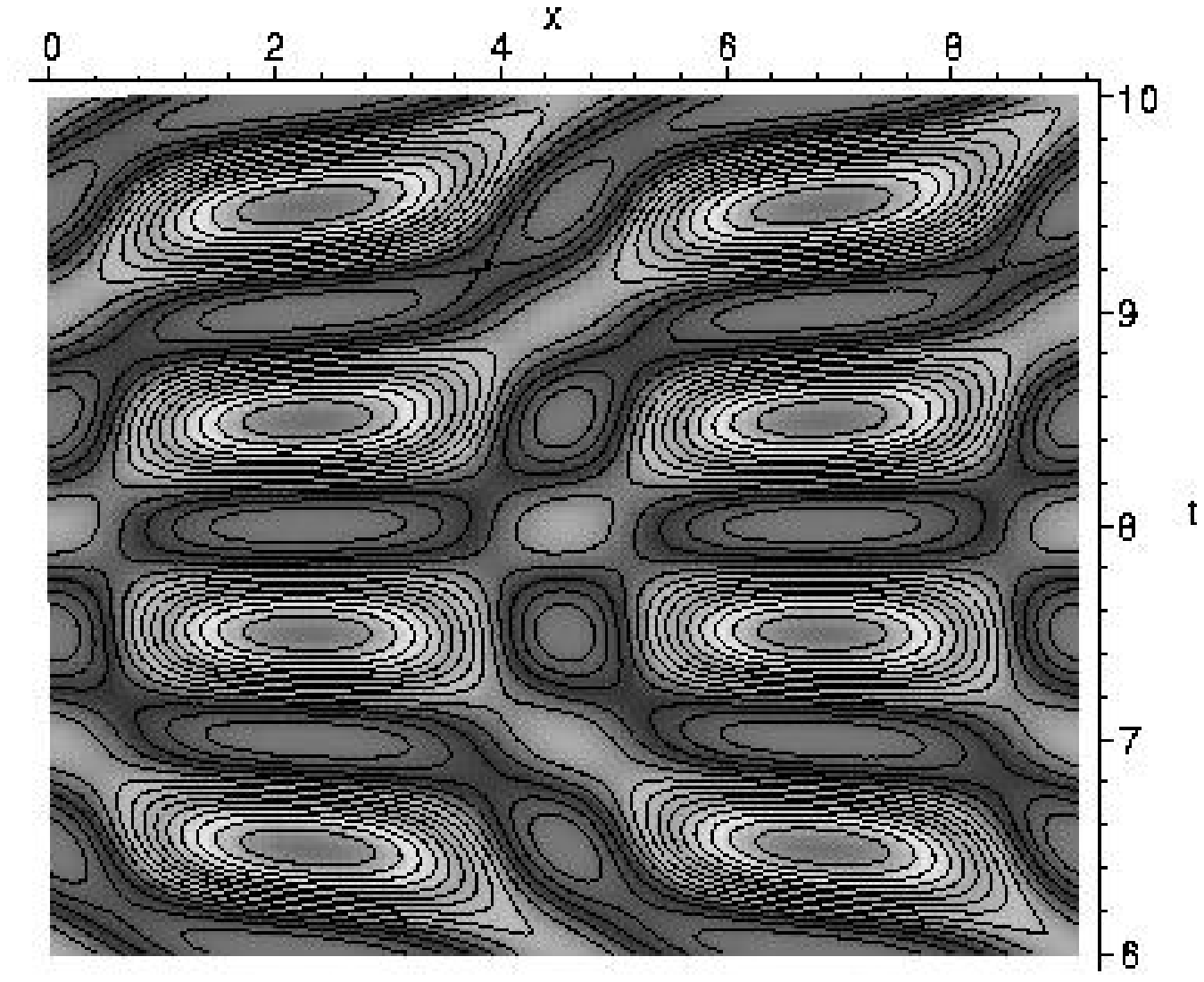}
\caption{
Top panel: 
two spatial periods of the time-periodic oscillations of
the density $n_2$ for the trigonometric solution in the OL potential.
Here $g=1$, 
$\alpha=1$, 
$m=1$, 
$\omega_1=\omega_2=\frac{1}{2}m^2+B_1+B_2$, and $V_0=1$, $A_1=2$,
$A_2=V_0-A_1=1$, $B_1=1$, $B_2=2$. 
Bottom panel: two spatial periods of a 
quasiperiodic-in-time oscillation of 
the density $n_2$ 
for the elliptic function solution given in the text.
Here $m=1$, 
$g=1$, $\alpha=\pi$,
$\omega_1=\frac{1}{2}m^2(1+k^2)-A_2$, $\omega_2=\frac{1}{2}m^2-A_2$. 
Also $V_0=1$, $A_1=2$, $A_2=V_0+m^2 k^2-A_1=1$.
} 
\label{Fig0OL}
\end{figure}

Exact solutions in the presence of a potential are not often available
in explicit form for Eq. (\ref{eqn:nls}). However, in the presence
of the physically relevant, optical lattice potential \cite{OL} of the form
$V=-V_0 \sin^2(mx)$ (in one spatial dimension), large classes of such 
stationary solutions can exist such as \cite{decon,bronski}
$\psi_{01}=(\sqrt{B_1}\cos(mx)-i
\sqrt{B_1+A_1}\sin(mx))e^{-i
\omega_1 t}$, $\psi_{02}=(\sqrt{B_2}\cos(mx)-i
\sqrt{B_2+A_2}\sin(mx))e^{-i \omega_2 t}$, where
$\omega_{1,2}, A_{1,2}, B_{1,2}$ are appropriate constants. 
These solutions then become
exact, genuinely time-periodic solutions of Eq. (\ref{eqn:stark})
according to Eq. (\ref{unit}). Such a solution is given in the
top panel of Fig. \ref{Fig0OL}. The bottom panel of the figure illustrates
a quasi-periodic solution of Eq. (\ref{eqn:stark}), constructed from
a periodic one of the form $\psi_{01}=\sqrt{A_1}~\sn(mx,k)e^{-i
\omega_1 t}$, $\psi_{02}=\sqrt{-A_2}~\cn(mx,k)e^{-i \omega_2 t}$
in the elliptic function potential $V=-V_0~\sn^2(mx,k)$, which 
degenerates into the OL one for $k \rightarrow 0$.

\begin{figure}
    \includegraphics[width=3.2in]{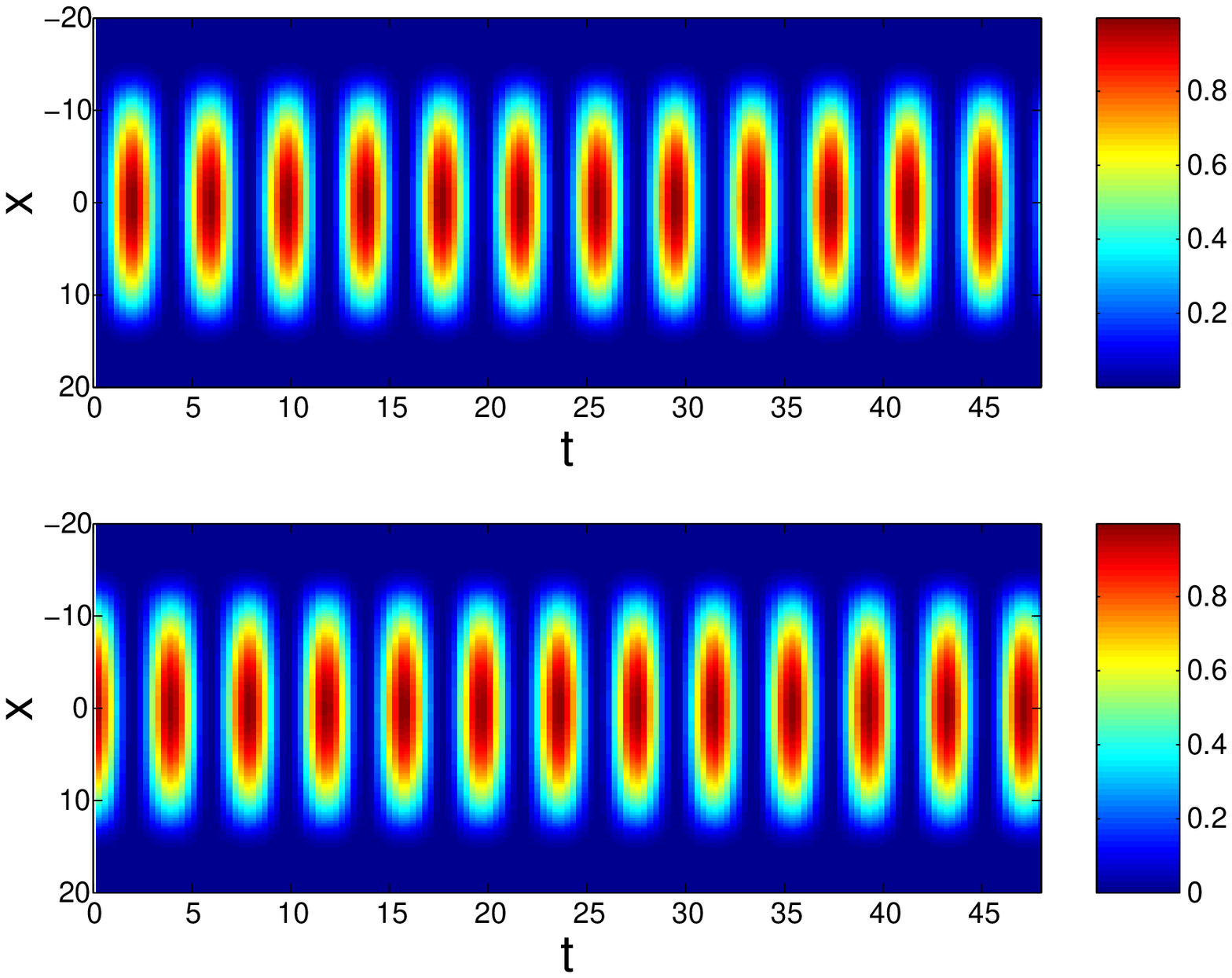} 
    \includegraphics[width=3.2in]{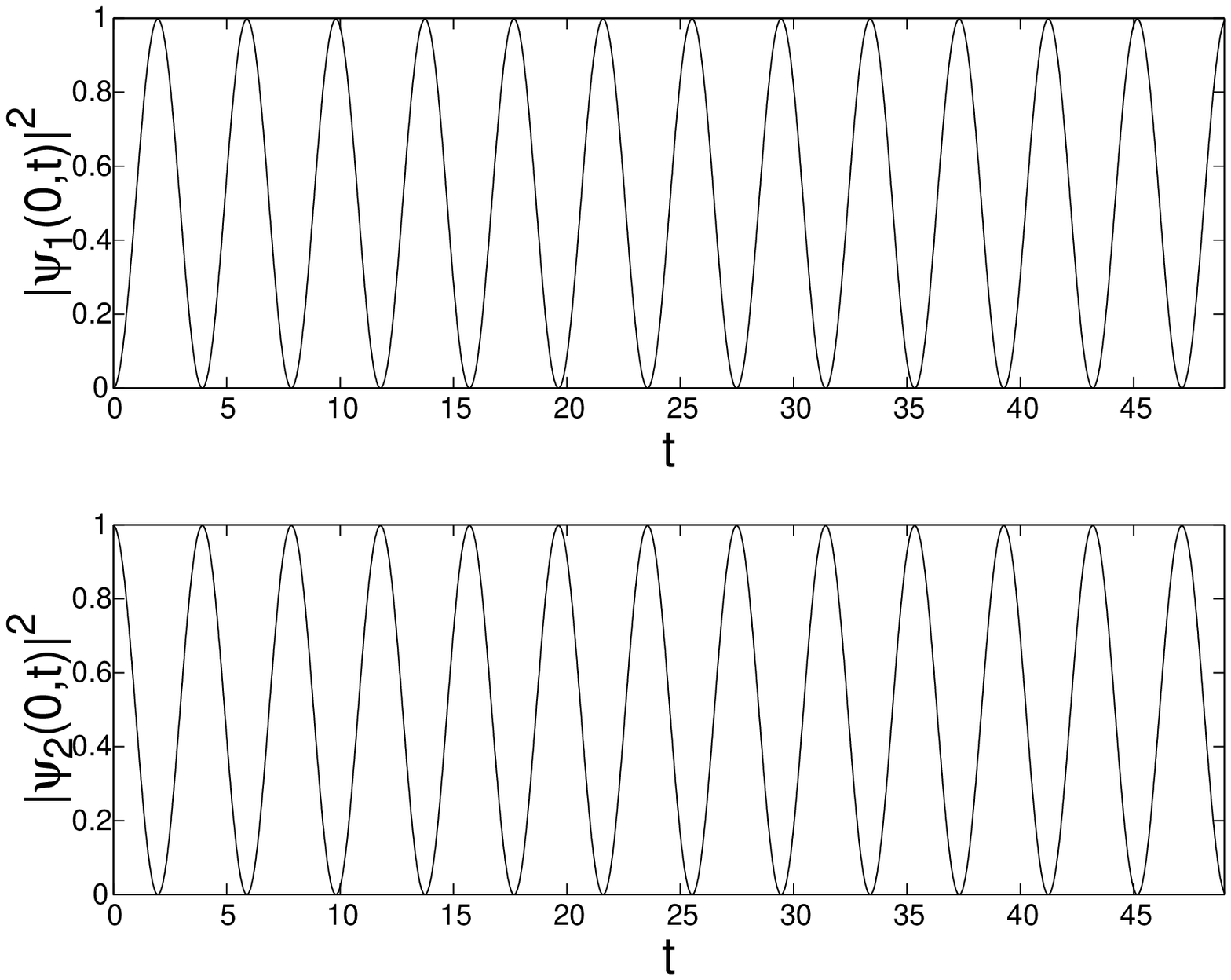} 
    \includegraphics[width=3.2in]{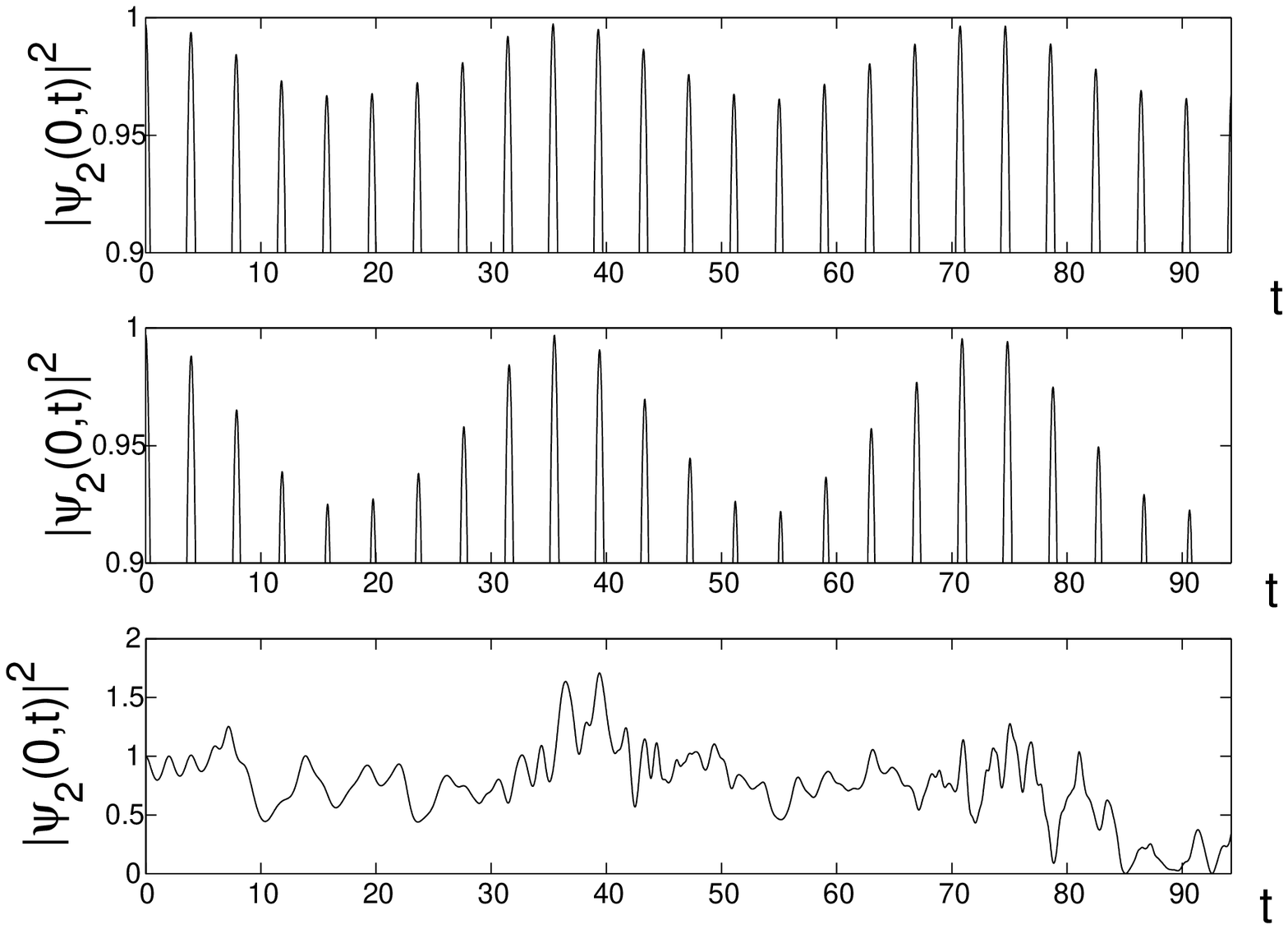}  
\caption{(Color Online)
Top two panels: the two subplots show {\it exact} Rabi oscillations
of a ground state BEC in the presence of a magnetic trap. The oscillation frequency is given by $\alpha=0.8$. The magnetic potential has a frequency $\Omega=0.1$. 
The two subplots show the contour of the space-time evolution of the density of
each component. 
Middle two panels: evolution of the density at the center of 
the magnetic trap i.e., at $x=0$ for the two components.
Bottom three panels: evolution at the center of the magnetic trap, 
for $h \neq g$, namely for 
$h=1.2$ (top subplot), $h=1.5$ (middle subplot) and $h=2$ (bottom
subplot).}
\label{Fig0MT}
\end{figure}

However, typically such explicit solutions are not known, e.g., in the
presence of a magnetic trap potential. Then one can use relaxational
methods such as a Newton iteration or imaginary time integration to
obtain stationary solutions of Eq. (\ref{eqn:nls}) and exploit the
time-dependence inherent in Eq. (\ref{unit}) to excite {\it exact}
matter wave oscillations between the two components, producing 
exact, non-stationary solutions of Eq. (\ref{eqn:stark}). A one
dimensional example of this strategy is shown in Fig. \ref{Fig0MT}
for the ground state BEC in a magnetic trap potential \cite{review,reviewus} 
of the form $V(x)=\Omega^2 x^2/2$. $\psi_2$ is initialized at
an exact stationary solution (in the presence of $V(x)$, obtained via
a Newton method), while $\psi_1(x,0)=0$. However, an interesting
question then concerns the potential persistence of the 
Rabi oscillations when the unitary transformation is
no longer exact (i.e., for $h \neq g$). This is examined in the
bottom set of panels in Fig. \ref{Fig0MT}.
Clearly, small deviations from the $h=g$ limit (relevant to the BEC context 
where  
$a_{11} \approx a_{22} \approx a_{12}$) lead to persistence of the matter 
wave oscillations but with a quasi-periodic beating character. 
Beyond a critical threshold, however, the oscillations
disappear and give rise to chaotic behavior.

\begin{figure}[tbp]
\includegraphics[width=3.2in]{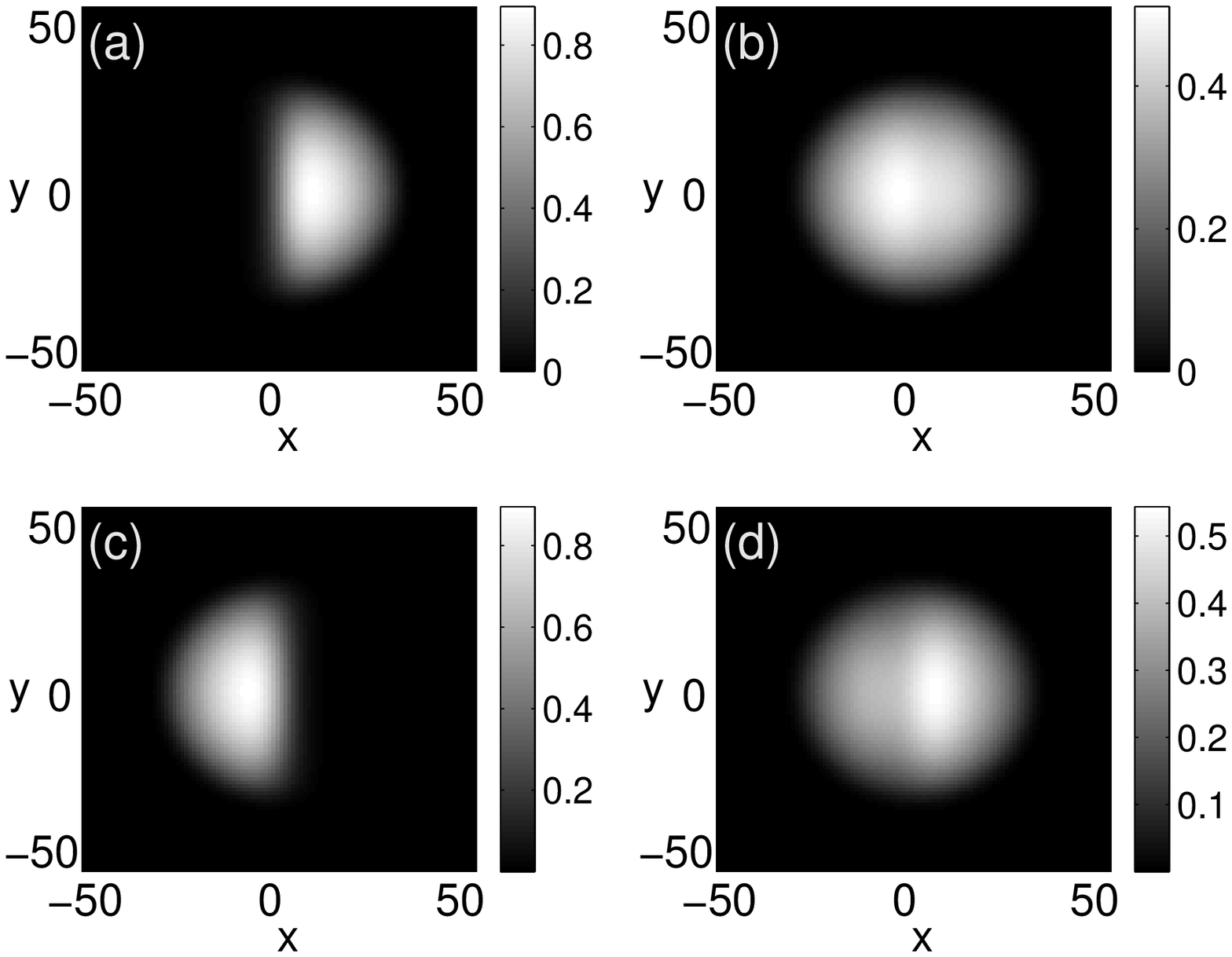}
\includegraphics[width=3.2in]{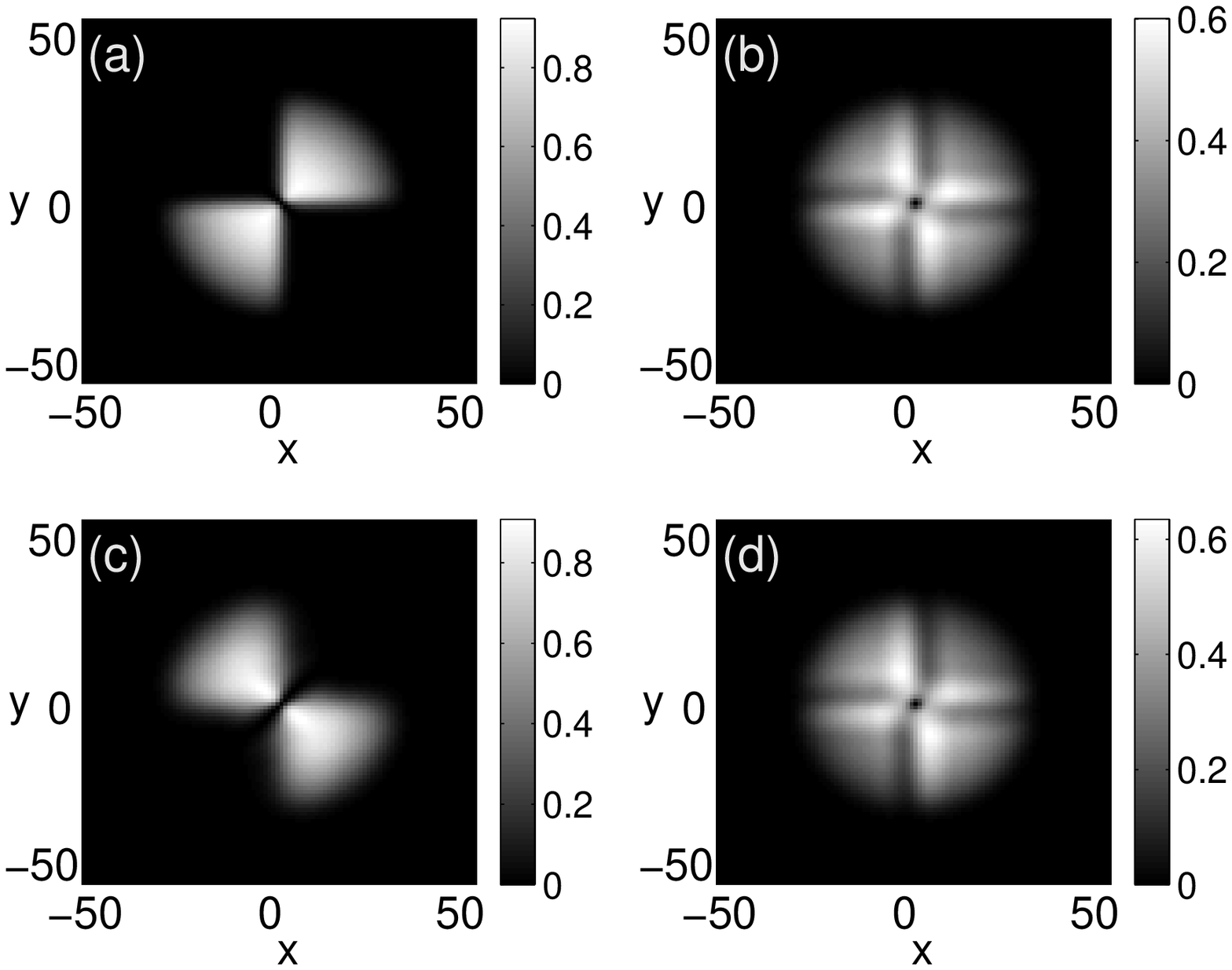}
\caption{
Top panels: Contour plots of the density $|\psi_{1}|^{2}$ for a dipolar 
DW (the density $|\psi_{2}|^{2}$ of the other species is complementary to 
$|\psi_{1}|^{2}$) for $t=0$ (a), $T/4$ (b), $T/2$ (c), and $3T/4$ (d), 
with $T=\pi/\alpha \approx 15.7$ ($\alpha=0.2$); $\Omega=0.045$, $\Delta=-0.061$. The pattern persists for long times with the 
species ``interchanging places''.
Bottom panels: Same as the top but for the 
quadrupolar DW with $\Delta=-9 \times 10^{-4}$ (the respective 
nonlinearity coefficients pertain to $^{87}$Rb). The Rabi 
oscillation of this sloshing DW gives the impression of a rotating propeller.}
\label{Fig1}
\end{figure}

Naturally, the same idea for constructing nonstationary solutions can be
carried over to two spatial dimensions. We illustrate the principle in the case
of domain-walls (DWs) and vortices in two spatial dimensions (relevant
results, but for a rotating trap were reported in \cite{sp}). As concerns the
DW solutions, instead of using the more ``standard'' circular DWs (between a
less repulsive component in the middle of the trap and a more repulsive one
forming an outer shell) \cite{shenoy,esry}, we will use the recently proposed 
dipolar (i.e., rectilinear) and quadrupolar (i.e., cross-like) DWs of
\cite{propeller}. 
Once such a stationary state is reached for Eq. (\ref{eqn:nls})
(via imaginary time integration),
using it as initial condition in 
Eqs. (\ref{lceqa})-(\ref{lceqb}) produces a time periodic solution of the 
latter. Such examples are shown in Fig.  \ref{Fig1} for the different 
types of DWs in the case of a magnetic trap 
potential $V(r)=(1/2)\Omega^{2}r^{2}$ ($r^{2}\equiv x^{2}+y^{2}$). 
We should note that the DW can, in principle, exist if the 
immiscibility condition $\Delta \equiv a_{11}a_{22}-a_{12}^{2} \le 0$ 
is satisfied. 
In particular, the quadrupolar DWs 
can persist even for $\Delta=0$ (or $g=h=1$) and hence the 
resulting ``rotating propeller'' solutions are exact time periodic solutions 
of Eq. (\ref{eqn:stark}). In the bottom panel of Fig. \ref{Fig1}, a DW cross 
is shown for the practical case of $a_{11}=1.03$, $a_{12}=1$ and $a_{22}=0.97$ 
(these values pertain to two different spin states of $^{87}$Rb \cite{dsh}). 
On the other hand, the dipolar DWs only persist if $\Delta \le -0.061$, 
which cannot be satisfied for the $^{87}$Rb parameters. In the top panel 
of Fig. \ref{Fig1} such a dipolar DW is shown, in the limiting case 
$\Delta=-0.061$ (for $g=1$ and $h=1.03$); apparently, the time-periodic, 
dipolar DWs are only approximate solutions. Similarly to the 1D results, 
if $h \neq g$ (or, generally, $\Delta \neq 0$), we have found that there 
is a maximum (minimum) critical $h_{c}$ ($\Delta_{c}$) up to which the Rabi 
oscillations persist [for the bulk of each component]: $h_{c}=1.67$ 
($\Delta_{c}=-1.78$) for dipolar DWs and $h_{c}=1.28$ ($\Delta_{c}=-0.64$) 
for the quadrupolar DWs. 

\begin{figure}[tbp]
\includegraphics[width=3.2in]{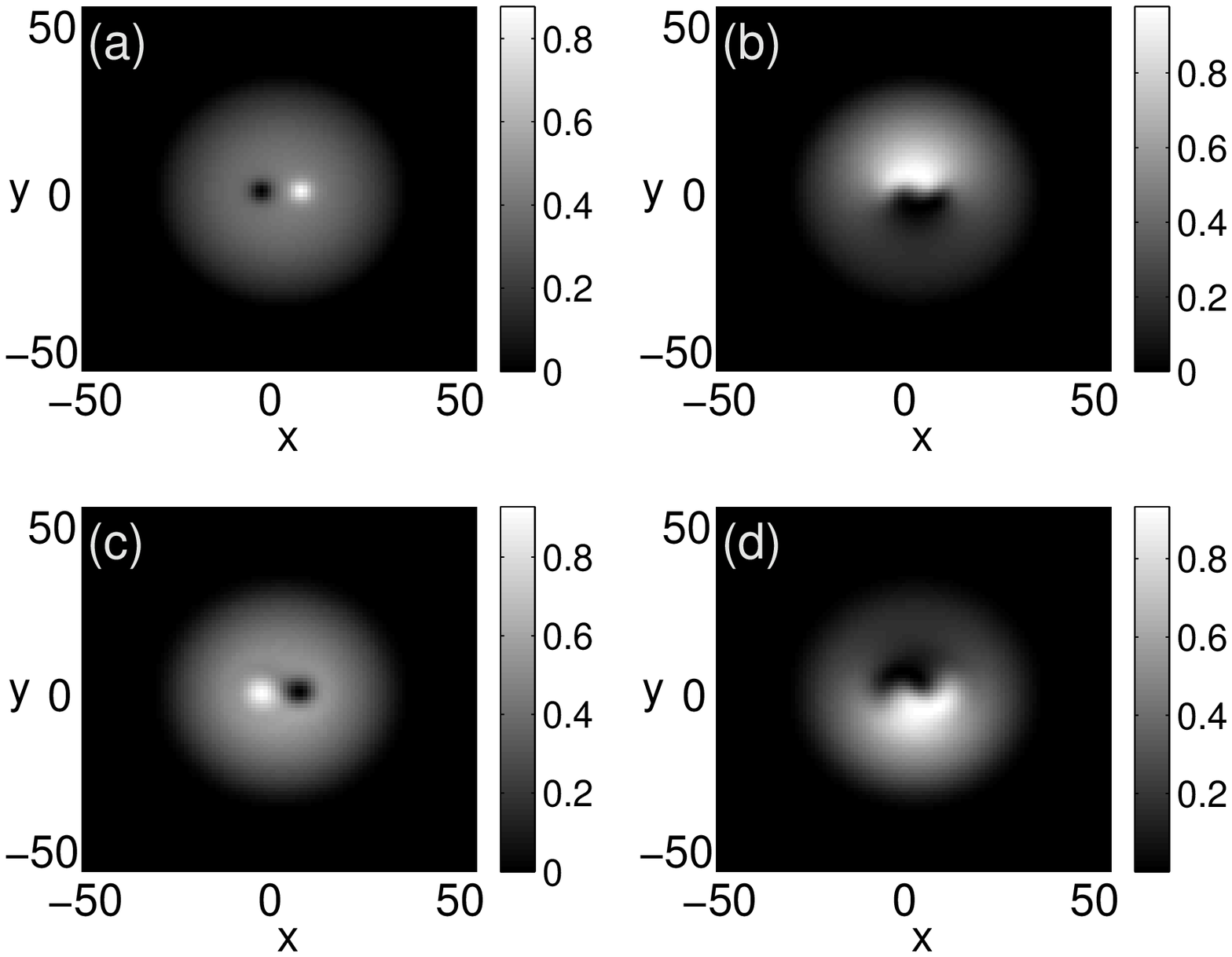}
\includegraphics[width=3.2in]{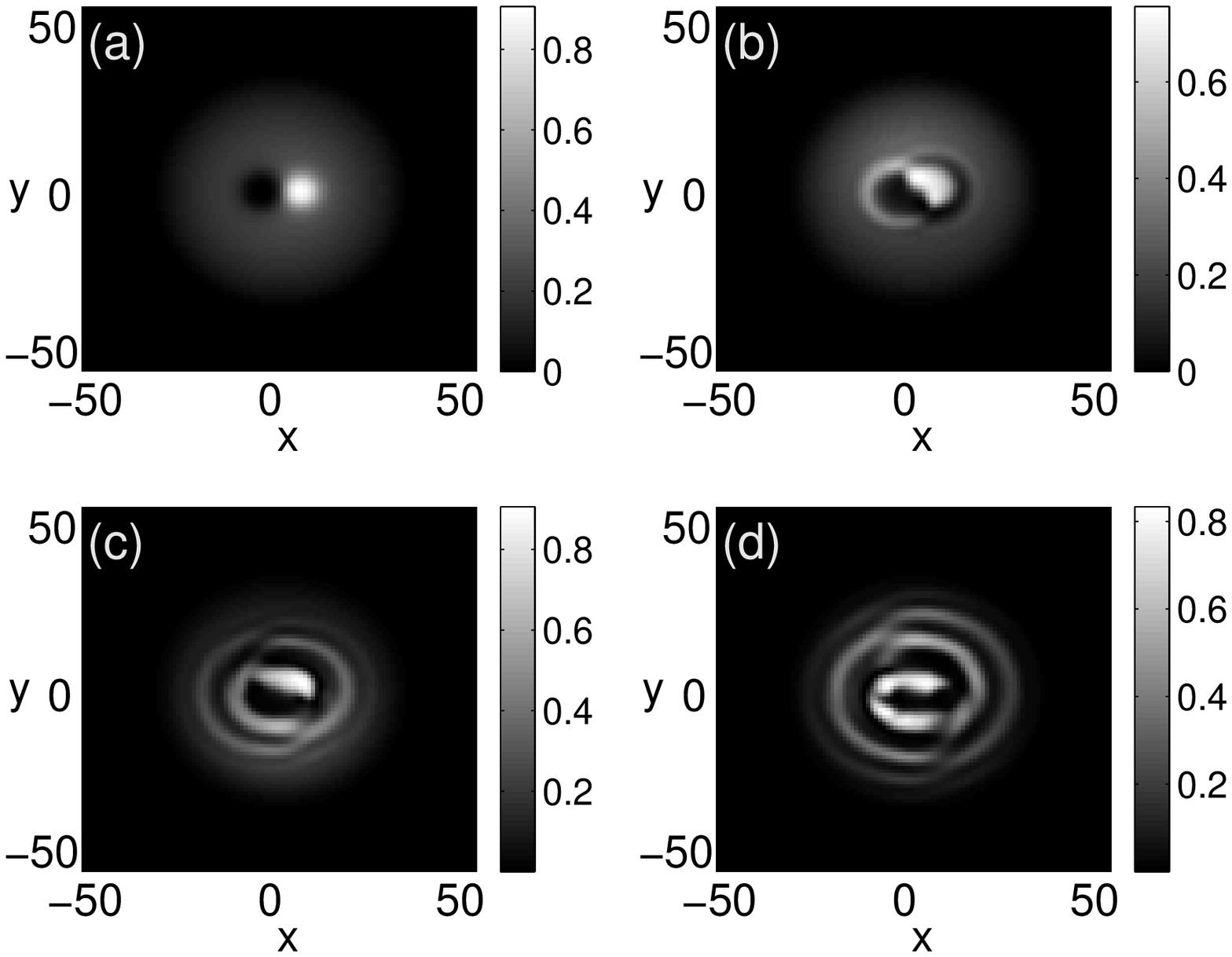}
\caption{
Top panels: Contour plots of $|\psi_{1}|^{2}$ for two coupled 
vortices, initially placed at $x= \pm 5$, for $t=0$ (a), $T/4$ (b), 
$T/2$ (c), and $3T/4$ (d), 
with $T=\pi/ \alpha \approx 15.7$ ($\alpha=0.2$);
$\Omega=0.045$, $\Delta=-9 \times 10^{-4}$ ($^{87}$Rb). The vortices ``interchange locations'' (in a structure resembling a spiral wave).
Bottom panels: Same as the top but for $\Delta=-3$ ($g=1$, $h=2$). The configuration breaks up forming spiral patterns.}
\label{Fig2}
\end{figure}

Finally, one can use a similar construction for a pair of vortex structures 
with
two components (see, e.g., Fig. \ref{Fig2}). 
In this case, we initialize the imaginary time integration,
in the absence of the linear coupling, with one component having a
vortex centered at $(5,0)$, while the other has a vortex at $(-5,0)$. After
the configuration relaxes to the stationary vortex pair solution of 
Eqs. (\ref{lceqa})-(\ref{lceqb}), we again turn on the linear coupling and obtain a spiral rotation between the vortices resembling a spiral wave. In this case also, there is
a critical $h_{c}=1.32$ ($\Delta_{c}=-0.74$) beyond which the regularity of Rabi oscillations is destroyed. In this case, the breakup leads to the formation of spiral patterns in the condensate.

{\it Conclusions and Generalizations.} In this Letter, we have
illustrated the possibility of coupled Bose-Einstein condensates
to sustain {\it exact} periodic and quasi-periodic patterns in the presence of an experimentally realizable, linear coupling between the components. 
Similar results should be immediately applicable to linearly coupled optical systems close to the Manakov limit. A unitary transformation, commuting with the nonlinear kernel was identified as the source of such solutions and as a way of ``factoring out'' the linear coupling by means of time-dependent oscillatory behavior with a frequency equal to the strength of the linear coupling. We demonstrated the relevance of this transformation in constructing various solutions in the presence of external potentials such as the optical lattice and the magnetic trap potential. We also illustrated the robustness of the mechanism in demonstrating that the phenomenon persists even for a wide range of parameter values (rather than only for the special yet experimentally relevant case of equal inter- and intra-species interaction for which it is exact). 

Finally, we would like to indicate that this mechanism is not restricted to the particular case of two linearly coupled components, but in fact generalizes to higher numbers of components (e.g., 3 linearly coupled hyperfine states can also be realized in the context of BECs \cite{porter}). In this case, the term $\alpha P \bpsi$ in Eq. (\ref{eqn:stark}) should be substituted with $\hat{\alpha}\bpsi$, where $\hat{\alpha}$ is a symmetric $n \times n$ matrix (with zeros along the diagonal). The unitary transformation then becomes $U(t)=e^{-i t\hat{\alpha}}$; e.g., in the special case 
where $n=3$ and all off-diagonal elements are identical,
then $U(t)=[\beta_{ij}]$, with diagonal elements 
$\beta_{jj}=(1/3)(2e^{i \alpha t}+e^{-2i \alpha t})$ and 
off-diagonal ones $\beta_{ij}=\beta_{jj}+e^{i \alpha t}$.
 


{\bf Acknowledgements} This work was supported in part by NSF-DMS-0139093
(BD), NSF-DMS-0204585, NSF-CAREER, and the
Eppley Foundation for Research (PGK).

\hspace{-30mm}


\end{multicols}

\end{document}